\documentstyle [12pt]{article}
\begin{document}

\setcounter{page}{1}
\title{Does the QGP fire ball really exist ?}
\author{G.A.Kozlov\\
{\em Bogoliubov Laboratory of Theoretical Physics}\\
\em Joint Institute for Nuclear Research\\
\em 141980 Dubna, Moscow Region, Russia}
\date{}
\maketitle
\begin{abstract}

{\small The strongly interacting matter under extreme
conditions of temperature for the conjectured deconfined phase of the
quark-gluon plasma is investigated. A systematic study of the form of the
thermal
ratio of the disorder deviation (TRDD) is presented with the emphasis on
the degree of deviation of the TRDD-function from
unity. The space-time evolution of quarks and gluons is studied in the
framework of two-particle correlation functions to predict the space-time
size of the closed deconfined phase.\\

PACS 12.38.Mh }
\end{abstract}
\section{Introduction}
\ \ \ \ \  In the near future a lot of efforts will be needed in searching
for the
quark-gluon plasma (QGP). The existence of a deconfined phase of gluons and
quarks has been predicted by Quantum Chromodynamics (QCD) [1].
Heavy-ion experiments at the CERN Large Hadron Collider (LHC)
with the program ALICE [2] will start
next century, aimed at  possible experimental confirmation
of the hypothetical theoretical predictions of QGP. It is supposed that
in heavy-ion collisions the hadron matter occurs as a strong thermal squeezed
state for a short time period. A deconfined phase, QGP, is possible
at the energy density $\sim 2.4~GeV/fm^{3}$ in this thermalized matter [3].
However, up to now there are no fixed and guaranteed signals which would
make it possible to identify the occurrence of QGP.
Matsui and Satz [4] pointed out that the
signature of the phase transition from the nuclear matter to QGP
would be the reduction of heavy vector meson yield. There is a suppression
of a heavy charmonium state $J/\Psi$ in heavy
ion collisions via the Debye color screening effect in a special mode phase
of quarks and gluons-the deconfined phase. It is pointed out that charmed
quarks and antiquarks would leave the zone of deconfinement before forming a
 hadron composed of charmed quarks.

    Following the fruitfull idea of Matsui and Satz [4] we suppose that\\
- at multi-TeV ($\sim 5-10~TeV$) collider energies the gluon luminosity is
far larger than the quark luminosity for charmed or beauty hybrid
intermediate states, where their constituents are distorted
temporarily by the deconfined environment during the short time scale just
after the fusion of two gluons coming from colliding ions;\\
- the charmed or beauty hybrid mesons turn into the s-wave charmonium or
bottomonium and $\gamma$ quantum.\\
Here, in the naive picture, the deconfined state of QGP is
occupied by an arbitrary number of gluons and/or quarks (g/q). It is
clear that there are no interactions between gluons and quarks for a short
time $\tau$ if the time scale for the QGP existence $\tau=\tau_{QGP}$ is
larger than the characteristic time scale $\tau_{V}$ ($\tau_{QGP}>
\tau_{V}$) to form the quark bound state with the average radius $<r^{2}>^
{1/2}$=2$\tau_{V}$/$m_{q}$ for $m_{q}$ quark mass. Otherwise, there are no
unbounded quarks if $\tau_{QGP}< 0.32 fm$ and $0.2 fm$ for $J/\Psi$ and
$\Upsilon$-particles, respectively. The radiative decay of charmed hybrid
mesons into $J/\Psi$ with the decay width $\sim 4.0 \alpha_{s}$ keV has been
estimated in [5].
There is a very popular point of view in literature that a (de)confinement
phase transition is predicted to occur at the typical energy scale involved,
the temperature $T_{c}\sim$ QCD scale $\Lambda_{QCD}\sim$ 200 MeV.
This critical
temperature $T_{c}$ is close to the limiting one in hadron interactions
firstly indicated by Hagedorn [6]. The only remaining problem which
has been transfering
>from one paper to another for a long time is the puzzle of rough equality
of $T_{c}$ to other three fundamental quantities in QCD: the scale
$\Lambda_{QCD}$, the pion mass, and the current mass of a strange quark.

  But, among the issues related to QGP, we attract attention to the
problem of deconfined phase through the calculation of correlation and
distribution functions [7] in the thermal theory of quantized fields.
  In this paper, we consider the semiphenomenological model for the
QGP existence. To do this, we have to use the standard theory of
quantized fields replacing:\\
1. the asymptotic field operators and\\
2. the vacuum expectation values\\
by\\
1. the thermal field operators and\\
2. the thermal statistical averages,\\
respectively, in order to formulate correlation and distribution functions of
produced particles.

    We assume that the heavy-ion collision (ex., Pb-Pb [2]) produces a
thermalized
quark-gluon gas because of the two-gluon fusion in the local chemical and
thermodynamic equilibrium. The hot as well as dense gas of particles will
expand into the surrounding vacuum, thereby cooling down, and the particles
will
interplay with each other and dilute untill free gluons and/or quarks,
g/q, as
well as particles composed of quarks leave this equilibrium. Corrections
to the free two-particle correlation functions (CF) taking account of the
interaction of g/q with their source medium are very actual. This allows one
to get information about the geometry of the emitter-source. For
convenience, one can distinguish the fase of g/q in the equilibrium from
the free particles by the freeze-out hypersurface.

   In this paper, we would like to propose new features of formulation
withing the framework of the Langevin-type equation. The method of
Langevin equation and its extensions to the quantal case have been
suggested and considered in papers [8-10] and [11-13], respectively.
 We propose that rather complicated real physical processes to happen
in the QGP formation should be replaced by a one-constituent
(ex., gluon or quark) propagation provided by a special
kernel operator (in the evolution equation) to be considered as an
input of the model and disturbed by the random force $F$.
We assume $F$ to be the external source  proposed as
both a c-number function and an operator. Our evolution equation is an
operator one, so that there appear new additional problems about the
commutation relations and the ordering of operators, which do not exist in
the classical case [13].

   Based on the thermal operator-field technique, in Sec. 2 we introduce a
thermal ratio of the disorder deviation (TRDD),
reflecting the degree of deviation, from unity, of the ratio of the
two-particle
thermal momentum-dependent distribution to two one-particle thermal
distribution functions of produced particles, gluons and/or quarks in a partly
deconfined phase state. We study the
four-momentum correlations of two identical particles, g/q, which can be both
useful and instructive in heavy ion collisions to infer the shape of the
particle emitter-source (Sec.3). The sensitivity of the TRDD-functions to
the size of the emitter will be given in Sec. 3.
Within these features, the canonical formalism in a
stationary state in the thermal equilibrium (SSTE) is formulated, and a closed
structural resemblance between the SSTE and standard quantum field theory
is revealed.\\
\section{Correlation and distribution functions}
   To clarify the internal structure of the disordering of particles
produced in heavy ion collisions, we have to use the consistent approach
based on the evolution of dynamical variables as well as the extension to
different modes provided by virtual transitions.

   Let us consider a hypothetical system of the quark-gluon excited local
thermal phase in QCD where a canonical operator $a(\vec{k},t)$ and
its Hermitian conjugate $a^{+}(\vec{k},t)$  occur. In the naive representation
, we suppose that in this phase there are independent particle sources
localized at the points $x_{\mu}$ and all the particles do not interact after
their emission.
We formulate the
distribution functions (DF) of produced particles (gluons and quarks) in
terms of
point-to-point equal time temperature-dependent thermal CF of two operators
$$w(\vec{k},\vec{k}^{\prime},t;T)=\langle a^{+}(\vec{k},t
)\,a(\vec{k}^{\prime},t)\rangle = $$
$$= Tr [a^{+}(\vec{k},t)\,a(\vec{k}^{\prime},t)e^{-H\beta}]/Tr
 (e^{-H\beta})\ .$$
 Here, $\langle ...\rangle$ means the procedure of thermal statistical
averaging; $\vec{k}$ and $t$ are, respectively, momentum and time variables,
$e^{-H\beta}/Tr(e^{-H\beta})$ stands for the standard density operator in the
equilibrium and the Hamiltonian $H$ is given by the squared form of the
annihilation $a_{p}$ and creation $a_{p}^{+}$ operators for bose- and fermi-
particles,
$H=\sum_{p}\epsilon_{p}a^{+}_{p}a_{p}$ (the energy $\epsilon_{p}$ and
operators $a_{p}$, $a_{p}^{+}$ carry some index $p$ [14], where
$p_{\alpha}=2\pi\ n_{\alpha}/L, n_{\alpha}=0,\pm 1,\pm 2, ...; V=L^{3}$ is the
volume of the system considered); $\beta$
is the inverse temperature of the environment, $\beta=1/T$. We assume that
$w(\vec{k},\vec{k}^\prime,t;T)=0$ if $\vert\vec{k}-\vec{k}^{\prime}\vert <\
\mu_{k}$ for the characteristic thermolized massive scale $\mu_{k}$, where
the temperature of the environment T does not change.
  One can suppose that the real physical process is devided into two stages:\\
1. Thermolization of the hadron matter system or even the quark-gluon
formation because of the heavy-ion collision. There is an expansion of the
thermolized space-time as well as a freeze-out process.\\
2. Upon cooling to the critical temperature, $T_{c}\sim 200^{o}C$, the
thermolized hadron system or the phase with free g/q is decaying into
secondary particles which could be observed. The first stage,
characterizing free particles in SSTE, allows one to make any field
operator using its expansion like (ex., for bose particles)
$$\Phi_{b}(x_{\mu})=\varphi(x_{\mu})+\varphi^{+}(x_{\mu})\ ,$$
where
\begin{eqnarray}
\label{e1}
\varphi(x_{\mu})=\int d^{3}\vec{k}\ v_{k}\ a(\vec{k},t)\ ,
v_{k}=\frac{e^{i\vec{k}\vec{x}}}{[(2\pi)^{3}\ 2\ \Delta(\vec{k})]^{1/2}}\ ,
\end{eqnarray}
and $\Delta(\vec{k})$ is the element of the invariant phase volume which will
be defined later.
     The standard canonical commutation relation (CCR)
\begin{eqnarray}
\label{e2}
{\left\lbrack a(\vec{k},t),a^{+}(\vec{k}^{\prime},t)
\right\rbrack}_{\pm}=\delta^{3}(\vec{k}-\vec{k}^{\prime})
\end{eqnarray}
at every time t is used as usual for bose (-) and fermi (+)-operators.
The next step is to introduce the TRDD-function $D(\vec{k},\vec{k}^{\prime},t
)=R(\vec{k},\vec{k}^{\prime},t)-1$ reflecting the deviation from unity
of the R-ratio
\begin{eqnarray}
\label{e3}
R(\vec{k},\vec{k}^{\prime},t)=W(\vec{k},\vec{k}^{\prime},t)/[W(\vec{k},t)
\cdot W(\vec{k}^{\prime},t)]\ .
\end{eqnarray}
The $R$-function (~\ref{e3}~) is defined as the probability
to find two particles, gluons or quarks, with momenta $\vec{k}$ and $\vec{k}^
{\prime}$ in the same event at the time $t$ normalized to the single
spectrum of these particles. Here, $W(\vec{k},t)$ stands for the one-particle
thermal DF
\begin{eqnarray}
\label{e4}
W(\vec{k},t)=\langle b^{+}(\vec{k},t)\ b(\vec{k},t)\rangle\ ,
\end{eqnarray}
\begin{eqnarray}
\label{e5}
b(\vec{k},t)=a(\vec{k},t)+\phi(\vec{k},t)
\end{eqnarray}
of the particles emitted (gluons or quarks) under an assumption of
occurrence of the random source-function $\phi(\vec{k},t)$ being an operator,
in general. The two-particle DF $W(\vec{k},\vec{k}^{\prime},t)$ looks like
\begin{eqnarray}
\label{e6}
W(\vec{k},\vec{k}^{\prime},t)=\langle b^{+}(\vec{k},t)\ b^{+}(\vec{k}
^\prime,t)\ b(\vec{k},t)\ b(\vec{k}^{\prime},t)\rangle\ .
\end{eqnarray}
This would allow one to estimate the possibility to find any constituents
(gluons and quarks) in the excited QCD matter. In order to be applied to
future experimental search for QGP, it should be important to compare
the measured TRDD function D for heavy-ion collisions at the TeV scale with
those calculated in a spherically symmetric model which describes the
transverse momentum spectra of g/q. Based on the theoretical model considered
here one can calculate and give the prediction for the space-time size of the
deconfined system.
   As a starting point, let us consider, for simplicity, the random
source-function $\phi(\vec{k},t)$ in (~\ref{e5}~) as a c-number one. The
R-ratio (~\ref{e3}~) of the DF (~\ref{e6}~) and (~\ref{e4}~) can easily
be calculated with the result leading to
$$R_{b}(\vec{k},\vec{k}^{\prime},t)=1+\frac{\sigma_{b}(\vec{k},\vec{k}^\prime,t
)}{\sum_{i=4}^{4}\alpha_{i}} $$
for bose-particles, while for fermi-ones we obtain
$$R_{f}(\vec{k},\vec{k}^\prime,t)=R_{b}(\vec{k},\vec{k}^{\prime},t)-\frac
{2\ \alpha_{1}}{\sum_{i=1}^{4}\alpha_{i}}$$
under the condition $\sigma_{b}(\vec{k},\vec{k}^\prime,t)>2\alpha_{1}$, where
$$\sigma_{b}(\vec{k},\vec{k}^\prime,t)=w(\vec{k},\vec{k}^\prime,t)\cdot
w(\vec{k}^\prime,\vec{k},t)+w(\vec{k},\vec{k}^\prime,t)\cdot\phi^{+}(\vec{k}
^\prime;t)\phi(\vec{k},t)+\ $$
$$+w(\vec{k}^\prime,\vec{k},t)\cdot\phi^{+}(\vec{k},
t)\phi(\vec{k}^\prime,t)\ ,$$
$$\alpha_{1}=w(\vec{k},\vec{k})\cdot w(\vec{k}^\prime,\vec{k}^\prime)\ ,
\alpha_{2}=w(\vec{k},\vec{k})\cdot{\vert\phi(\vec{k}^\prime)\vert}^2\ ,$$
$$\alpha_{3}=w(\vec{k}^\prime,\vec{k}^\prime)\cdot{\vert\phi(\vec{k})\vert}
^2\ ,\alpha_{4}={\vert\phi(\vec{k})\vert}^2\cdot {\vert\phi(\vec{k}^\prime)
\vert}^2\ .$$
Thus, the deviation from unity, reflecting the naive result, is defined
by the positive TRDD-functions $D_{b/f}$ ranging from 0 to 1:
$$D_{b}=\frac{\sigma_{b}}{\sum_{i=1}^{4}\alpha_{i}}\ ,\ \ \ \
D_{f}=\frac{\sigma_{b}-2\ \alpha_{1}}{\sum_{i=1}^{4}\alpha_{i}}\ .$$
The complete correlation is provided by the condition $D_{b/f}$=1, while
$D_{b/f}$=0 corresponds to the naive mode with zeroth correlation.

  The formal structure of the $D_{b/f}$- functions does not allow one to
reveal the origin of the particle disordering. We have indicated the formal
scheme only in order to obtain a nontrivial result ($D\neq 0$) to reveal
the correlation effect with the random source represented by the c-number
function $\phi(\vec{k};t)$. To avoid the formal representation of the
result and to do some translation into an operator space, one has to go into
some evolution scheme for the set of all operators to be used.

   One of our aims is to give a useful tool reflecting the evolution
properties of propagating particles in a randomly distributed environment.
 Let us consider an elementary particle "moving" with the
momentum $\vec{k}$ in the quantum equilibrium phase space under the influence
of the random force coming from surrounding particles. The evolution
equations
for these particles described in terms of the operators $b(\vec{k},t)$
and $b^{+}(\vec{k};t)$ are
\begin{eqnarray}
\label{e7}
i\ \partial_{t}b(\vec{k},t)+A(\vec{k},t)=F(\vec{k},t)+P\ .
\end{eqnarray}

\begin{eqnarray}
\label{e8}
-i\ \partial_{t}b^{+}(\vec{k},t)+A^{*}(\vec{k},t)=F^{+}(\vec{k},t)+P\ .
\end{eqnarray}
One can identify both $b$ and $b^{+}$ with the special mode operators of
the quark and gluon fields or their combinations [15] having the dependence
on the thermolized QCD matter fields occurring due to heavy-ion collision.
In the classical picture, equations (~\ref{e7}~) and (~\ref{e8}~) are just
the Langevin-type ones for the Brownian motion of particles. Here,
$P$ and $F(\vec{k},t)$
stand for the stationary external force and the random one, respectively,
both acting
>from the environment. The only operator $F$ has a zeroth value of the
statistical average, $\langle{F}\rangle=0$.
The correlation between two operators $F(\vec{k},t)$ represents the
fluctuation dissipation theorem of the second kind [16].
The interaction of the particles
considered with the surrounding ones as well as providing the propagation
is given by the operator $A(\vec{k},t)$ which can be defined as the one
closely related to the dissipation force. A most simple form of the operator
$A(\vec{k},t)$ should be the following:
\begin{eqnarray}
\label{e9}
A(\vec{k},t)=\int_{-\infty}^{+\infty}K(\vec{k},t-\tau)\ b(\vec{k},\tau)\ d\tau\ .
\end{eqnarray}
 Here, an interplay of quarks and gluons with
surrounding particles is embedded into the interaction complex kernel
$K(\vec{k},t)$, while the real physical transitions are provided by the
random source operator $F(\vec{k},t)$ ( see eq. (~\ref{e7}~)).
The random evolution field operator $K(\vec{k},t)$ in (~\ref{e14}~) stands for
the random noise and it is assumed to vary stochastically with a $\delta$-
like equal time correlation function
\begin{eqnarray}
\label{e10}
\langle K^{+}(\vec{k},\tau)\ K(\vec{k}^\prime,\tau)\rangle =2\
{(\pi\alpha)}^{1/2}\ \kappa\ \delta (\vec{k}-\vec{k}^\prime)\ .
\end{eqnarray}
In the case of the large value of the correlation scale squared $\alpha$
one can get
\begin{eqnarray}
\label{e11}
\langle K^{+}(\vec{k},\tau)\ K(\vec{k}^\prime,\tau)\rangle\rightarrow
\kappa\  \exp[-\vec{z}^2/(4\alpha)]\ \ \ \ \  as\ \ \  \alpha\rightarrow\infty\ ,
\end{eqnarray}
where $\vec{z}^2={(\vec{k}-\vec{k}^\prime)}^2$ while $\kappa$ is the strength
of the
noise characterized by the Gaussian distribution function $exp[-\vec{z}^2/
(4\alpha
)]$. Hence, both $\kappa$ and $\alpha$ in (~\ref{e10}~) and (~\ref{e11}~)
define the effect of the Gaussian
noise on the evolution of g/q in the thermolized environment.

  The formal solution of (~\ref{e7}~) in the operator form
in $S(\Re_{4})$ ($k^{\mu}=(\omega=k^0,k_{j}))$ is
\begin{eqnarray}
\label{e12}
\tilde b(k_{\mu})=\tilde a(k_{\mu})+\tilde\phi(k_{\mu})\ ,
\end{eqnarray}
where the operator $\tilde a(k_{\mu})$ is expressed via the Fourier
transformed operator $\tilde F(k_{\mu})$ and the Fourier transformed kernel
function $\tilde K(k_{\mu})$ (coming from (~\ref{e9}~) as
\begin{eqnarray}
\label{e13}
\tilde a(k_{\mu})=\tilde F(k_{\mu})\cdot [\tilde K(k_{\mu})-\omega]^{-1}\ ,
\end{eqnarray}
while the function $\tilde\phi(k_{\mu})$ is provided by the function
$\sim P\cdot [\tilde K(k_{\mu})-\omega]^{-1}$.
In order to obtain the solution (~\ref{e12}~), we have used the fact that at
large $t$ the terms in the Fourier transformed evolution equation in
$S(\Re_{4})$, containing the fast oscillating factors, weakly tend to zero
as $ t\rightarrow +\infty$ as well as $t\rightarrow -\infty$. The random
force operator $F(\vec{k},t)$ can be expanded by using the Fourier integral
$$F(\vec{k},t)=\int_{-\infty}^{+\infty}\frac{d\omega}{2\pi}\ \psi(k_{\mu})\
\hat c(k_{\mu})\ e^{-i\omega t}\ ,$$
where the form $\psi(k_{\mu})\cdot\hat c(k_{\mu})$ is just the Fourier
operator $\tilde F(k_{\mu})=\psi(k_{\mu})\cdot\hat c(k_{\mu})$ and the
canonical operator $\hat c(k_{\mu})$ obeys the commutation relation
$${\left\lbrack\hat c(k_{\mu}), \hat c^{+}(k_{\mu}^{\prime})\right\rbrack}_{\pm}=
\delta^{4}(k_{\mu}-k^{\prime}_{\mu})\ .$$
The function $\psi(k_{\mu})$ is determined by the condition
\begin{eqnarray}
\label{e14}
\Delta(\vec{k})=\int_{-\infty}^{+\infty}\frac{d\omega\cdot\omega}{2\pi}
{\left [\frac{\psi(k_{\mu})}{\tilde K(k_{\mu})-\omega}\right ]}^{2}\ ,
\end{eqnarray}
where $\Delta(\vec{k})$ comes from the field expansion (~\ref{e1}~), and the
CCR (~\ref{e2}~) is used. The requirement $\Delta(\vec{k})={(\vec{k}^2+m^2)}
^{1/2}$ in (~\ref{e14}) immediately leads to the condition (see also [15])
$$\int_{-\infty}^{+\infty}\frac{d\omega}{2\pi}
{\left [\frac{\psi(k_{\mu})}{\tilde K(k_{\mu})-\omega}\right ]}^{2}=1\ . $$

    The time correlation of the random force operator is defined by
$$\langle F^{+}(\vec{k},t)\ F(\vec{k},t+\tau)\rangle=\int_{-\infty}^{+\infty}
\int_{-\infty}^{+\infty}\frac{d\omega}{2\pi}\frac{d\omega^\prime}{2\pi}
e^{it(\omega-\omega^\prime)}e^{-\omega^\prime\tau} $$
$$\times\langle\psi^{+}(\omega,\vec{k})\ \hat c^{+}(\omega,\vec{k})\
\psi(\omega^\prime,\vec{k})\ \hat c(\omega^\prime,\vec{k})\rangle\ .$$
In the case when the constituents (gluons or quarks) are "moving" inside
the closed system being in the equilibrium state, the magnitude of the initial
time $t$ does not have any priority compared to other initial time points.
Hence, the time correlation $\rho(\vec{k},t)=\langle F^{+}(\vec{k},t)\ F(\vec
{k},t+\tau)\rangle$ should not carry the dependence of the time variable.
This fact can be realized by putting
$$\langle\psi^{+}(\omega,\vec{k})\hat c^{+}(\omega,\vec{k})\psi(\omega^
\prime,\vec{k})\hat c(\omega^\prime,\vec{k})\rangle=
2\pi\langle\hat c^{+}(k_{\mu})\hat c(k_{\mu})\rangle{\vert\psi(k_{\mu})\vert}^2\delta
(\omega-\omega^\prime)\ .$$
Thus, the time correlation is translated into the following one:
$$\rho(\vec{k},t)=\int_{-\infty}^{+\infty}\frac{d\omega}{2\pi}\ {\vert\psi
(k_{\mu})\vert }^2\ \langle\hat c^{+}(k_{\mu})\ \hat c(k_{\mu})\rangle\
e^{-i\omega\tau}\ .$$
In fact, the quantity
${\vert\psi(k_{\mu})\vert}^2\langle\hat c^{+}(k_{\mu})\hat c(k_{\mu})\rangle $
is the spectral function characterizing the random
force operator $\tilde F(k_{\mu})$.

\section{Distribution functions in $S(\Re_{4})$ and space-time size.}
   In order to study the enhanced probability for emission of two identical
particles, g/q, and since we are interested in the finite size of the QGP
formation, the useful information comes from the consideration of DF
in $S(\Re_{4})$. We introduce the ratio $R$ of DF as follows:
\begin{eqnarray}
\label{e15}
R_{b/f}(k_{\mu},k_{\mu}^\prime;T)=\frac{\tilde W(k_{\mu},k_{\mu}^\prime;T)}
{\tilde W(k_{\mu})\cdot\tilde W(k_{\mu}^\prime)}\ ,
\end{eqnarray}
where $\tilde W(k_{\mu},k_{\mu}^\prime;T)$ stands for two-particle
probability density subject to Bose/Fermi symmetrization and defined as
$$\tilde W(k_{\mu},k_{\mu}^\prime;T)=\langle\tilde b^{+}(k_{\mu})\ \tilde b^
{+}(k_{\mu}^\prime)\ \tilde b(k_{\mu})\ \tilde b(k_{\mu}^\prime)\rangle\ ,$$
while
$$\tilde W(k_{\mu})=\langle\tilde b^{+}(k_{\mu})\ \tilde b(k_{\mu})\rangle\ $$
is related to the single-particle quantity for a particle with momentum
$k_{\mu}$. Using the Fourier solution of equation (~\ref{e7}~) in $S(\Re_{
4})$, one can get R-ratios for DF obeying to Bose-
\begin{eqnarray}
\label{e16}
R_{b}(k_{\mu},k_{\mu}^\prime;T)=1+D_{b}(k_{\mu},k_{\mu}^\prime;T)
\end{eqnarray}
and Fermi-particles
\begin{eqnarray}
\label{e17}
R_{f}(k_{\mu},k_{\mu}^\prime;T)=R_{b}(k_{\mu},k_{\mu}^\prime;T)-
2\ \frac{\Xi(k_{\mu})\cdot\Xi(k_{\mu}^\prime)}{\tilde W(k_{\mu})
\cdot\tilde W(k_{\mu}^\prime)}
\end{eqnarray}
where
\begin{eqnarray}
\label{e18}
D_{b}(k_{\mu},k_{\mu}^\prime)=\frac{\Xi(k_{\mu},k_{\mu}^\prime)[\Xi(k_{\mu}^
\prime,k_{\mu})+\tilde\phi^{+}(k_{\mu}^\prime)\tilde\phi(k_{\mu})]+
\Xi(k_{\mu}
^\prime,k_{\mu})\tilde\phi^{+}(k_{\mu})\tilde\phi(k_{\mu}^\prime)}{
\tilde W(k_{\mu})\cdot\tilde W(k_{\mu}^\prime)}.
\end{eqnarray}
In order to estimate the thermolized space-time size of QGP we need to
have the detailed knowledge of CF $\Xi(k_{\mu},k_{\mu}^\prime)$. Using
the general approach, one cannot do this estimation but suppose the
formal expansion to be applied to the function $\Xi(k_{\mu},k_{\mu}^\prime)$,
namely
$$\Xi(k_{\mu},k_{\mu}^\prime)=
\sum_{\zeta}\langle k^\prime\vert\zeta\rangle
 f(\omega;T)\langle\zeta\vert k\rangle\ ,$$
where the sum runs over all possible $\zeta$-states of the environment
while the $f$-function stands for the temperature-dependent one.
Using (~\ref{e13}~) and the Fourier
operator $\tilde F(k_{\mu})=\psi(k_{\mu})\cdot\hat c(k_{\mu})$
the two-particle CF $\Xi(k_{\mu},k_{\mu}^\prime)$ looks like
\begin{eqnarray}
\label{e19}
\Xi(k_{\mu},k_{\mu}^\prime)=\langle\tilde a^{+}(k_{\mu})\ \tilde a(k_{\mu}^
\prime)\rangle \cr
=\frac{\psi^{*}(k_{\mu})\cdot\psi(k_{\mu}^\prime)}{[\tilde K^{*}(k_{\mu})-
\omega]\cdot[\tilde K(k_{\mu}^\prime)-\omega^\prime]}\cdot\langle\hat c^{+}
(k_{\mu})\ \hat c(k_{\mu}^\prime)\rangle\ .
\end{eqnarray}
The definition of the thermal statistical average $\langle\hat c^{+}(k_{\mu})
\ \hat c(k_{\mu}^\prime)\rangle$ is related to the Kubo-Martin-Schwinger
(KMS) condition [17]
\begin{eqnarray}
\label{e20}
\langle a(\vec{k}^\prime,t^\prime)\ a^{+}(\vec{k},t)\rangle = \langle a^{+}
(\vec{k},t)\ a(\vec{k}^\prime,t-i\beta)\rangle\cdot \exp(-\beta\ \mu)\ ,
\end{eqnarray}
where $\mu$ is the chemical potential. Using the KMS condition (~\ref{e20}~)
and the Fourier transformed solution $\tilde a(k_{\mu})$ (~\ref{e13}~), one
can conclude
that the thermal statistical averages for the $\hat c(k_{\mu})$-operator
should be presented in the following form:
\begin{eqnarray}
\label{e21}
\langle\hat c^{+}(k_{\mu})\ \hat c(k_{\mu}^\prime)\rangle =\delta^{4}(k_{\mu}-
k_{\mu}^\prime)\cdot n(\omega,T)\ ,
\end{eqnarray}
\begin{eqnarray}
\label{e22}
\langle\hat c(k_{\mu})\ \hat c^{+}(k_{\mu}^\prime)\rangle =\delta^{4}(k_{\mu}-
k_{\mu}^\prime)\cdot [1\pm n(\omega,T)]
\end{eqnarray}
for Bose (+)- and Fermi (-)-statistics where $n(\omega,T)=\{\exp[(\omega-\mu)
\beta]\pm 1\}^{-1}$. Inserting CF (~\ref{e19}~) into
(~\ref{e18}~) and taking into account that the
$\delta^{4}(k_{\mu}-k_{\mu}^\prime)$-function should be changed by the smooth
sharp
function $\Omega(r)\cdot \exp(-q^2/2)$, one can get the following expression
for the $D_{b}$-function
\begin{eqnarray}
\label{e23}
D_{b}(k_{\mu},k_{\mu}^\prime;T)= \lambda(k_
{\mu},k_{\mu}^\prime;T)\ \exp(-q^2/2) \cr
\times [n(\bar{\omega},T)\Omega(r) \exp(
-q^2/2)+\tilde\phi^{*}(k_{\mu}^\prime)\tilde\phi(k_{\mu})+\tilde\phi^{*}(k_{
\mu})\tilde\phi(k_{\mu}^\prime)]\ ,
\end{eqnarray}
where
$$\lambda(k_{\mu},k_{\mu}^\prime;T)=\frac{\Omega(r)}{\tilde W(k_{\mu})\cdot
\tilde W(k_{\mu}^\prime)}\cdot n(\bar{\omega},T)\ ,\ \bar{\omega}=\frac{1}{2}
(\omega+\omega^\prime)\ .$$
The function $\Omega(r)\cdot n(\omega;T)\cdot \exp(-q^2/2)$ describes the
space-time size of the QGP fire-ball. Choosing the z-axis along the
two-heavy-ion collision axis one can put
$$q^2=(r_{0}\cdot Q_{0})^2+(r_{z}\cdot Q_{z})^2+
(r_{t}\cdot Q_{t})^2\ , $$
$$Q_{\mu}=(k-k^\prime)_{\mu}, Q_{0}=\epsilon_
{\vec{k}}-\epsilon_{\vec{k}^\prime}, Q_{z}=k_{z}-k_{z}^\prime, Q_{t}=
{{[(k_{x}-k_{x}^\prime)}^2+{(k_{y}-k_{y}^\prime)}^2]}^{1/2}\ , $$
$$\Omega(r)\sim r_{0}\cdot r_{z}\cdot r_{t}^2\ ,$$
where $r_{0}$, $r_{z}$ and $r_{t}$ are time-like, longitudinal and transverse
"size" components of the QGP fire-ball. For estimation the 4-dimensional
structure of the space-time correlators (~\ref{e21}~), (~\ref{e22}~) we
have to identify $r_{0}$, $r_{z}$ and $r_{t}$, resp., with the averages of
the "time" component $\tau$, z and $\sqrt {x^{2}+y^{2}}$ on the freeze-out
hypersurface with the temperature T.
Formally, the function $D_{b}$ (~\ref{e23}~) is the positive one ranging from
0 to 1. Since there are large values of $\sqrt {Q_{\mu}^2}=\sqrt
{{(k_{\mu}-k_{\mu}^\prime)}^2}$ in (~\ref{e23}~), all the
$k_{\mu}(k_{\mu}^\prime)$ dependence
should be changed by the average value $\bar k_{\mu}=0.5 {(k+k^\prime)}_{\mu}$.
The quantitative information (longitudinal $r_{z}$ and transverse $r_{t}$
components of the QGP spherical volume, the temperature T of the environment)
could be extracted by fitting the theoretical formula (~\ref{e23}~) to
the measured TRDD function and estimating the errors of the fit parameters.
Formula (~\ref{e23}~) indicates that a chaotic g/q source emanating from the
thermolized g/q fireball exists. Hence, the measurement of the space-time
evolution of the g/q source would provide information of the g/q emission
process and the general reaction mechanism. In formula (~\ref{e23}~) for the
$D_{b}$- function, the temperature of the environment enters through the
two-particle CF $\Xi(k_{\mu},k_{\mu}^\prime;T)$. If T is unstable the
$R_{b/f}$-functions (~\ref{e15}~) will change due to a change of
DF $\tilde W$ which, in fact, can be considered as an effective density
of the g/q source. In the case when the temperature in the center of the
fire-ball is higher than that of the border, this leads to the largest
contribution to the g/q production. This fact could be "translated into" the
possible statement that $\bar k_{T}<<\bar k_{z}$.
   Formula (~\ref{e16}~) looks like the following expresion for the
experimental R-ratio using a source parametrization:
\begin{eqnarray}
\label{e24}
R_{T}(r)=1+\lambda_{T}(r)\cdot \exp(-{r_{t}^2\cdot Q_{t}^2}/2 -{r_{z}^2
\cdot Q_{z}^2}/2)\ ,
\end{eqnarray}
where $r_{t}(r_{z})$ is the transverse (longitudinal) radius parameter of the
source with respect to the beam axis, $\lambda_{T}$ stands for the effective
intercept parameter (chaoticity parameter) which has a general dependence of
the mean momentum of the observed particle pair. Here, the dependence on the
source lifetime is omitted. Since $0<\lambda_{T}<1$, one can conclude that the
effective function $\lambda_{T}$ can be interpreted as a function of the
core particles to all particles produced. The chaocity parameter
$\lambda_{T}$ is the temperature-dependent and the positive one defined by
\begin{eqnarray}
\label{e25}
\lambda_{T}(r)=\frac{{\vert\Omega(r)\
n(\bar{\omega};T)\vert }^2}{\tilde W_{0}(k_{\mu},k_{\mu}^\prime)}\ ,
\end{eqnarray}
where $\tilde W(k_{\mu})\cdot\tilde W(k_{\mu}^\prime)$ is replaced by
$\tilde W_{0}(k_{\mu},k_{\mu}^\prime)$ for convenience regarding the
point of view that one can distinguish different particles.

   Comparing (~\ref{e18}~) and (~\ref{e23}~) one can identify
$$\Xi(k_{\mu},k_{\mu}^\prime)=\Omega(r)\cdot n(\bar{\omega};T)
\cdot \exp(-q^2/2)\ .$$
In the simple case, the random source function $\tilde\phi(k_{\mu})$ in
(~\ref{e18}~) should be the following [18]
\begin{eqnarray}
\label{e26}
\tilde\phi(k_{\mu})={[\alpha\cdot \Xi(k_{\mu})]}^{1/2}\ ,
\end{eqnarray}
where $\alpha$ is a real positive parameter. Neglecting
$\tilde\phi(k_{\mu})$-
source we immediately obtain the trivial result
$$D_{b}(k_{\mu},k_{\mu}^\prime;T)=\tilde\lambda (\bar{\omega};T)
\cdot \exp(-q^2)\ ,$$
where the temperature-dependent chaoticity $\tilde\lambda (\bar{\omega};T)$ is
$$\tilde\lambda (\bar{\omega};T)=\frac{n^{2}(\bar{\omega};T)}{n(\omega;T)
\cdot n(\omega^\prime;T)}\ .$$
Saving the random source-function we can find the formal representation for
the TRDD-function $D_{b}$ using the $\alpha$-factorization (~\ref{e26}~)
\begin{eqnarray}
\label{e27}
D_{b}(q^2;T)=\frac{\tilde\lambda^{1/2}(\bar{\omega};T)}{(1+\alpha)(1+
\alpha^\prime)}e^{-q^{2}/2}\left [\tilde\lambda^{1/2}(\bar{\omega};T)
e^{-q^{2}/2}+2{(\alpha\alpha^\prime)}^{1/2}\right]\ .
\end{eqnarray}
It is easily to see that in the vicinity of $q^{2}\approx 0$ one can get
the full correlation if $\alpha =\alpha^\prime =0$ and $\tilde\lambda (\bar
{\omega};T)$=1. Putting $\alpha =\alpha^\prime$ in (~\ref{e27}~) we find
the formal lower
bound on the space-time dimensionless size of the fire-ball for bose-system:
$$q^2_{b}\geq\ln\frac{\tilde\lambda (\bar{\omega};T)}{{[\sqrt{(\alpha+1)^{2}+
\alpha^2}-\alpha]}^2}\ .$$
In the case of fermi-particles, the following restriction on $q^{2}_{f}$ is
valid (see (~\ref{e17}~))
$$\ln\frac{\tilde\lambda (\bar{\omega};T)}{{[\sqrt{2\alpha(\alpha+1)+3}
-\alpha]}^2}\ \leq q^2_{f}\ \leq\ln\frac{\tilde\lambda (\bar{\omega};T)}
{{[\sqrt{\alpha^2+2}-\alpha]}^2}\ .$$

  In fact, the function $D_{b}(k_{\mu},k_{\mu}^\prime;T)$ in (~\ref{e23}~)
could not be observed in the heavy-ion experiment ALICE because of some
model uncertainties. In the standard consideration, the TRDD-function has
to contain
a background contribution as well as other physical particles (resonances)
which have not been included in the calculation of the $D_{b}$- function.
In order to be close to the experimental data, one has to expand the $D_{b}$-
function as projected on some well-defined function (in $S(\Re_{4})$)
of the relative momentum of two particles produced in heavy-ion collisions
$$D_{b}(k_{\mu},k_{\mu}^\prime;T)\rightarrow D_{b}(Q_{\mu}^{2};T)\ .$$
Thus, it will be very instructive to use the polynomial expansion which is
suitable to avoid any uncertainties as well as characterize the degree of
deviation from the Gaussian distribution, for example. In $(-\infty,+\infty)$,
a complete orthogonal set of functions can be obtained with the help of the
Hermite polynomials in the Hilbert space of the square integrable functions
with the measure $d\mu(z)=\exp(-z^2/2)dz$. The function $D_{b}$ corresponds to
this class if
$$\int_{-\infty}^{+\infty}dq \exp(-q^2/2)\ \vert D_{b}(q)\vert^n\ <\infty\ ,
 n=0,1,2,...\ .$$
The expansion in terms of the Hermite polynomials $H_{n}(q)$
\begin{eqnarray}
\label{e28}
D_{b}(q)=\lambda\sum_{n}\ c_{n}\cdot H_{n}(q)\cdot \exp(-q^2/2)\
\end{eqnarray}
is well suited for the study of possible deviation from both the experimental
shape and the exact theoretical form of the TRDD function $D_{b}$
(~\ref{e23}~). The coefficients $c_{n}$ in (~\ref{e28}~) are defined via
the integrals over
the expanded functions $D_{b}$ because of the orthogonality condition
$$\int_{-\infty}^{+\infty} H_{n}(x)\ H_{m}(x)\ \exp(-x^2/2)\ dx=\delta_{n,m}
\ .$$
Hence, the observation of the two-particle correlation (both for bose- and
fermi-symmetrization) enable to extract the properties of the structure of
$q^2$, i.e. the space-time size of QGP formation.\\
\section{Conclusion}
1. In this paper we investigated the finite temperature momentum correlations
(of two identical particles, gluons/quarks) which can be both useful and
instructive in multi-TeV heavy-ion collisions (e.g. Pb+Pb in ALICE
experimental program) to infer the shape
of the gluon/quark source-emitter. In fact, we have presented the method
of extracting the intercept and source parameters from the shape of the
TRDD-function.\\
2. We used the operator form evolution equation (~\ref{e7}~) and (~\ref{e8}~)
 for the special mode operators $b(\vec{k},t)$ and $b^{+}(\vec{k},t)$ (of
the gluon/quark field), respectively, in the thermolized equilibrium at the
freeze-out stage.\\
3. The relations between the CF $\Xi(k_{\mu},k_{\mu}^\prime)$ and the full
$R$-functions for bose (~\ref{e16}~)- and fermi (~\ref{e17}~)-particles at the
stage of the freeze-out are obtained. We have shown the sensitivity of the
correlation functions to the space-time geometry of the source-emitter (~\ref
{e23}~). In fact, the TRDD- function $D_{b}$ describes the size and shape
of the space-time domain where the secondary observed particles are
generated.\\
4. We can conclude that formally, the QGP size scale can be determined by the
evolution behavior of the field operators and the critical temperature
$T=T_{c}$ (see formula (~\ref{e23}~)).\\
5. In fact, the full $R$-ratio (~\ref{e15}~) is the function of four-momentum
difference $Q_{\mu}$ of two identical particles as well as the mean total
momentum $\bar{k}_{\mu}$. Since, the TRDD-function $D_{b}$ is the positive
one and restricted by 1, we expect that the $R$-ratio at too small values of
$Q_{\mu}$ starts from the fixed point $R(Q_{\mu}\rightarrow 0)=2-\epsilon$
$(\epsilon\rightarrow +0)$ and then falls down (with the Gaussian shape) up
to unity over some momentum scale interval of an order of the inverse source
size (see (~\ref{e24}~) and (~\ref{e25}~)).\\
6. A large momentum scale $Q_{\mu}$ determines the minimal size of the QGP
fire-ball volume for which $R(r_{\mu};T)$ deviates from its asymptotic value
of $R(r_{\mu}\rightarrow 0;T)=1$.\\
7. In order to use the quantitative information like longitudinal and
transverse source size, lifetime, we have to fit our expressions to the
measured $R$-functions.\\

\end{document}